\documentclass[preprint]{aastex}
\usepackage{ulem}

\slugcomment{}

\shorttitle{Eclipsing CV Lanning 386}

\shortauthors{Brady et al.}

\begin{document}
\title{The Eclipsing Cataclysmic Variable Lanning 386: 
Dwarf Nova, SW Sextantis Star, or Both?\footnote{Dedicated to Howard 
Lanning, who died in 2007 December.}}

\author{Steve Brady}
\email{sbrady10@verizon.net}
\affil{AAVSO, 25 Birch Street, Cambridge, MA 02138}

\author{John R. Thorstensen}
\email{john.thorstensen@dartmouth.edu}
\affil{Dartmouth College, Department of Physics and Astronomy, 6127 Wilder Lab,
Hanover, NH 03755}

\author{Michael D. Koppelman}
\email{michael@aps.umn.edu}
\affil{University of Minnesota, Department of Astronomy, 116 Church St. SE, Minneapolis, MN 55455}

\author{Jose Luis Prieto}
\email{prieto@astronomy.ohio-state.edu}
\affil{Ohio State University, Department of Astronomy, 140 W. 18th Ave., Columbus, OH 43210}

\author{Peter M. Garnavich, Alec Hirschauer, and Michael Florack}
\email{pgarnavi@nd.edu; ahirsch1@nd.edu; mflorack@nd.edu}
\affil{University of Notre Dame, Physics Department, 225 Nieuwland Hall, Notre Dame, IN 46556}

\begin{abstract}

We present photometry and spectroscopy of the suspected cataclysmic 
variable (CV) Lanning 386.  We confirm that it is a CV, and 
observe deep eclipses, from which we determine the 
orbital period $P_{\rm orb}$ to be $0.1640517 \pm 0.0000001$ d (= 3.94 h). 
Photometric monitoring over two observing seasons shows a 
very active system with frequent outbursts
of variable amplitude, up to $\sim 2$ mag.  The spectrum in quiescence is
typical of dwarf novae, but in its high state the system
shows strong \ion{He}{2} emission and a broad \ion{C}{4} Wolf-Rayet feature.
This is unusual for dwarf novae in outburst and indicates a high excitation.  
In its high state the system shows some features reminiscent of an 
SW Sextantis-type CV, but lacks others.  We discuss the classification
of this puzzling object.

%The \ion{He}{2} line is absent in the
%quiescent state. The Balmer and helium lines show a complex and time-varying
%relation between absorption and emission lines, suggesting that this is a new
%member of the SW~Sex class of CV.

\end{abstract}


%\section*{}
%\begin{quote}
%\textit{Dedicated to Howard Lanning, who died in 2007 December.
%}
%\end{quote}

\section{Introduction}

Cataclysmic variables (CVs) are close binary stars in which a
white dwarf primary accretes matter via Roche lobe overflow from
a secondary, which usually resembles a lower main-sequence star
\citep{warner}.
In most CVs an accretion disk encircles the white dwarf.  Optical
emission can come from several locations in the system.  The gas
stream from the secondary star strikes the outer rim of the
accretion disk producing a rapidly fluctuating bright spot (or
hot spot). The accretion disk often dominates the light, and
instabilities in the disk can produce large variations in the
system brightness.  Strong white-dwarf magnetic fields can
truncate or even entirely disrupt the accretion disk, allowing
material to fall directly onto the white dwarf.  In many systems
the accumulated hydrogen-rich material can undergo a
thermonuclear runaway, creating a classical nova outburst. The
range of phenomena seen in CVs means that determining the subclass
of any example can require extensive photometric and spectroscopic
study.

H. H. Lanning (\citealt{lanning1973} and subsequent papers)
conducted a search for ultraviolet-bright (UV) sources in the
galactic plane by visually examining plates taken with the Palomar
Oschin Schmidt.  To date, six supplements
have been published, revealing a total of 724 UV sources.  Six of
these are included in the {\it Catalog and Atlas of 
Cataclysmic Variables} \citep{downes2001} as confirmed
or suspected CVs; they are listed in Table \ref{lanningtable}.

\citet{eracleous2002} presented a spectrum of Lanning
386 which showed apparent CV features -- a blue continuum, 
He I absorption lines, and broad Balmer emission lines.
The extensive
photometric and spectroscopic observations we present here 
establish that
Lanning~386 is indeed a deeply-eclipsing CV with $P_{\rm orb} =
0.1640517 \pm 0.0000001$ d.  Intriguingly, it shares characteristics
of both the dwarf novae -- CVs that undergo occasional eruptions
thought to be caused by a disk instability -- and the SW Sextantis 
stars \citep{thorstensen1991}, a subclass of the persistently-bright
novalike variables.

\section{Observations}

\subsection{Photometry}

We began routine monitoring of Lanning
386 on 2005 June 25, with a 0.25 m robotic survey 
telescope at the lead author's
private observatory in southern New Hampshire.  The system was
programmed to automatically monitor a list of poorly-studied
CV candidates and to initiate time series photometry if 
and when an outburst was detected.  The first outburst of 
%Lanning 386 was detected 2005 September 2, after eleven
%  ... not consistent with date in table.
Lanning 386 was detected 2005 September 12 UT, after eleven
weeks of monitoring.  Time series observations (detailed
in Table \ref{photsumtable}) began 
immediately, and revealed two
eclipses separated by $\sim 3.9$ hr, about 1.5 mag deep.  
Thereafter, time series observations were obtained 
every clear night for 5 months in 2005 and 
another 5 months in 2006, yielding over
3,000 photometric observations.  
A 0.4 m telescope at the same observatory was used along
with the 0.25 m, and all observations were 
unfiltered due to the star's faintness.
Integration times were 300 sec with the 0.25 m telescope and 180
sec with the 0.4 m.  All images were bias subtracted and
flat-field corrected.  Ensemble photometry was performed using 12
relatively isolated comparison stars.
We established a rough zero point using $V$ magnitudes of four of
these stars (Figure \ref{finder_wide}) measured with the MDM 2.4m 
telescope (see below); Table \ref{cmptable} gives their positions
and magnitudes.
The mean scatter of the differential photometry
was $\sim 1$ percent.  The unfiltered photometry has a
significant color term, and the zero points of the two filter 
photometry sessions described below were slightly discrepant, 
so we estimate the uncertainty in the
photometric zero point to be $\pm 0.1$ mag.

Our filter photometry is from the 2.4 m Hiltner telescope at MDM
Observatory on Kitt Peak.  A $2048^2$ SITe CCD detector, 
cropped to $1024^2$ to minimize read time, gave a 4.7 arcmin
square field of view.  We obtained single sets of short 
(60 - 120 s) exposures through $UBVI$ filters on 2005 September 15 and
2007 June 24 UT, measured them using the 
IRAF\footnote[0]{IRAF is distributed by the National Optical Astronomy
Observatory, which is operated by the Association of Universities for
Research in Astronomy, Inc., under cooperative agreement with the National
Science Foundation.} implementation of 
DAOPHOT \citep{stetsondao}, and derived transformations to the 
Johnson/Cousins photometric system using observations of 
\citet{landolt92} standard-star fields.  The zero points for the
two sets of observations differed by $\sim 0.06$ mag in $V$.
The 2005 September 15 observations show
Lanning 386 at $V = 15.20$, $B-V = +0.08$, 
$U-B = -0.63$, and $V-I = +0.29$, while the 2007 June 24 observation
yields $V = 17.28$, $B-V = +0.27$, $U - B = -0.87$, and $V-I =
+1.07$.

\subsection{Spectroscopy}

Our spectra come from three different instruments on two different
telescopes, as listed in Table \ref{specjournal}.

Our most extensive spectroscopy is from the Hiltner telescope,
equipped with the MDM Modular spectrograph, a 600 line mm$^{-1}$
grating, and the same $2048^2$ SITe detector used for the filter
photometry.  The Modspec
observations were taken by one of us (JRT) without awareness that
eclipses had been discovered, and were therefore aimed largely at
determining $P_{\rm orb}$.
The spectra cover from 4200 to 7500 \AA\ , with 2 \AA\
pixel$^{-1}$ sampling and $\sim 3.5$ \AA\ resolution (FWHM).  The
observing and reduction procedures were as described in
\citet{sheets07}.  

We also obtained time-series spectra of Lanning~386 in its
bright state using the Kitt Peak 
National Observatory (KPNO) Mayall 4 m telescope and RC spectrograph.
The T2KB CCD detector, KPC10a grating, and a 1~arcsec 
slit provided an average dispersion of 2.8 \AA\ pixel$^{-1}$ and a useful 
range between 3700 and 8000 \AA , with some
order overlap beyond 7400 \AA . The slit was rotated to the parallactic angle to
prevent differential slit losses.
The spectra were bias subtracted, flat-field corrected using images of
an internal quartz lamp, extracted using the IRAF {\it twodspec} package,
and wavelength calibrated against HeNeAr lamp exposures taken from
time to time as the telescope tracked.  We applied
a flux calibration derived from observations
of the spectrophotometric standard BD +28$^\circ$ 4211.

Finally, we have two 600-s exposures of Lanning~386 taken during 
twilight on 2006 Dec 8 UT, using the MDM Hiltner telescope and
OSU Boller and Chivens CCD spectrograph (CCDS).  The spectra
were reduced in the same way as the KPNO data described 
above.  The 5000-6000 \AA\ continuum is $\sim 2.4$ mag fainter than 
in the KPNO spectra, confirming that the system was in quiescence.
  
%spectrograph (CCDS) beginning at 01:24 UT which corresponds to an
%orbital phase of 0.08, so the star was just coming out of eclipse at
%the start of the exposures. The spectra were reduced in the same way
%as the KPNO data and is also plotted in Figure \ref{spec1plot}. The
%continuum between 500 and 600~nm in the MDM spectrum was 2.4 mag
%fainter than the continuum of the KPNO spectrum at the same phase,
%confirming that the MDM spectrum was obtained during a quiescent
%phase.

\section{Analysis}

\subsection{Ephemeris}

%We investigated the periodic behavior of the system using Fourier analysis
%tools and the CLEANest algorithm \citep{foster1995}. Because this was an
%eclipsing system, with eclipses roughly $0.16^d$ apart, our search was
%conducted between $0.05^d < P < 0.33^d$ and produced a strong signal at $P =
%0.164059^{d} \pm 0.000004$ (Figure \ref{power}) and at harmonic frequencies.
Table \ref{eclipsestable} lists the observed times of mid-eclipse.  
These yield a unique ephemeris,
\begin{equation}\label{eqn1}
%\begin{array}{r@{}r@{}l}
%{\rm Min.~I = HJD~}2453625.58669 & {} + 0.1640517 & {} \times E  \\
%\pm 0.0004 & {} \pm 0.0000001, &  \\
%\end{array}
\hbox{Minimum light} = \hbox{Heliocentric JD } 2453625.5867(4) + 0.1640517(1) E,
\end{equation}
where $E$ is an integer and the parentheses give uncertainties in the 
last digits.
Radial velocities of the H$\alpha$ emission line (discussed below)
give essentially the same period, namely 0.1640530(11) d, without
ambiguity caused by uncertainty in the number of cycles elapsed
between observing runs.  All phases in this paper refer to the 
eclipse ephemeris.

% If it's not convincing, it is almost certainly not real -- JRT
%
%A search for shorter period signals showed a peak at $P=0.0279 \pm 0.0005$
%which might be due to quasi-periodic oscillations (QPOs) that are common in
%SW~Sex type CV \cite{patterson02}. The  semi-amplitude of the short period
%signal is $A=0.018\pm0.028$ but visual inspection of the data folded on this
%period is not convincing.  Future observations with higher photometric
%precision, finer time resolution and longer observing sessions would aid in
%the investigation of the high-frequency content of the light curve.
%
\subsection{Light Curve and Long-Term Variations}

\begin{figure}
\plotone{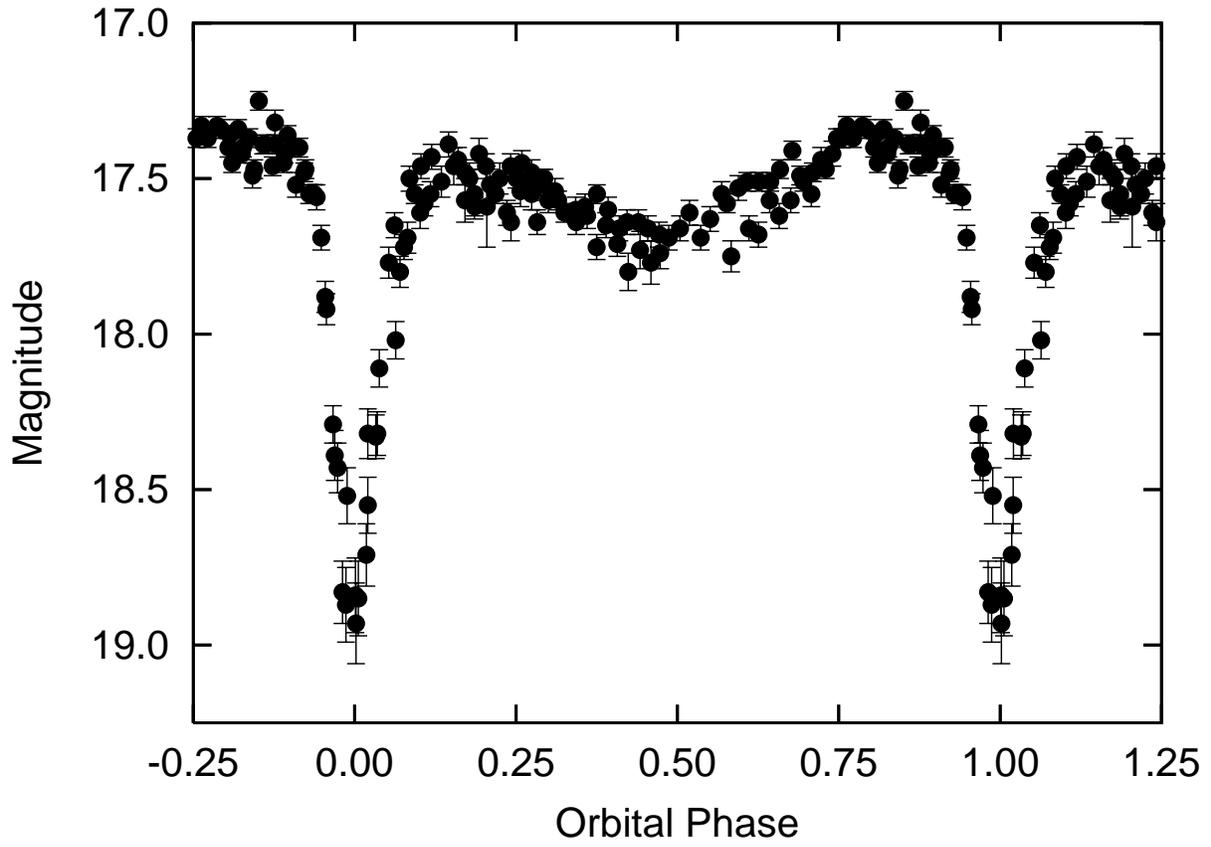}
\caption{Two nights of quiescent-state photometry folded on the orbital period. 
Note the prominent pre-eclipse hump. The observations are unfiltered, but calibrated 
to the Johnson $V$-band.}
\label{quietobplot}
\end{figure}

\begin{figure}
\plotone{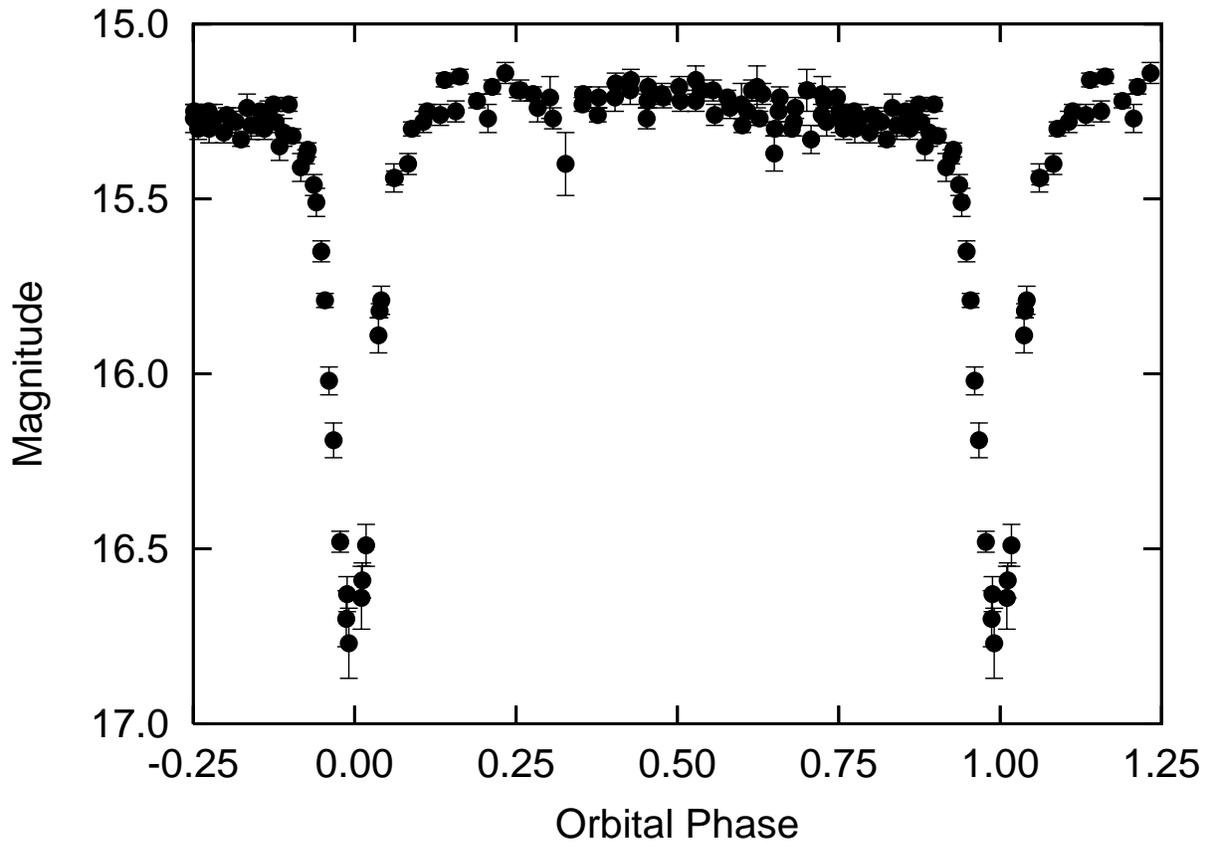}
\caption{Two nights in outburst folded on the orbital period.
The observations are unfiltered, but calibrated to the Johnson $V$-band.}
\label{outburstplot}
\end{figure}

In Figures \ref{quietobplot} and \ref{outburstplot} we show
representative lights curves in quiescence and outburst, constructed
by folding two consecutive observation sessions using the ephemeris
(Eqn.~\ref{eqn1}).  The quiescent plot, from observations made between
2005 September 27 and October 1 UT, shows a pre-eclipse hump; this
feature usually arises from the ``hot spot'' where gas streaming from
the secondary star strikes the disk.  The hot spot contribution is not
evident in the outburst plot, made from observations taken 2006 July
16 and 17 UT.  The eclipse depth is roughly the same in each state.

\begin{figure}
\plotone{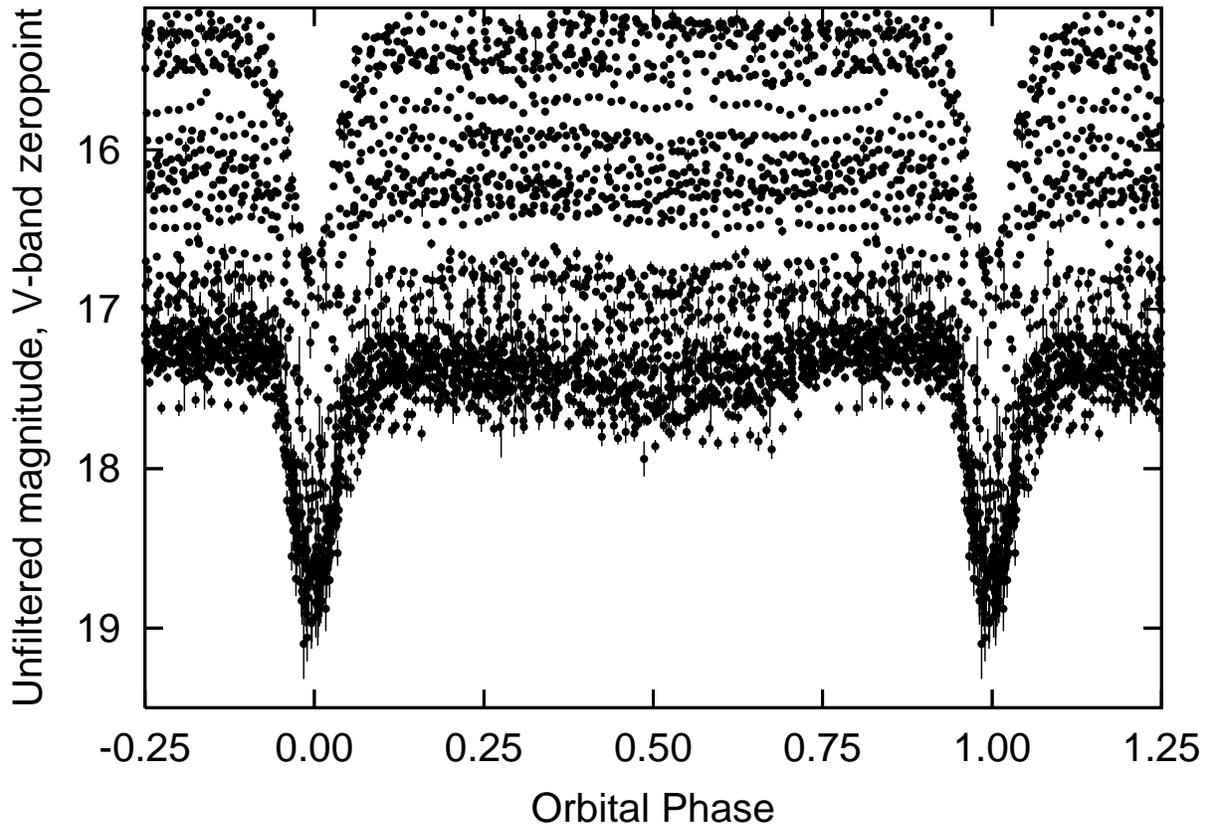}
\caption{All photometric observations folded on the orbital period. Note how the 
eclipse depth does not vary significantly with brightness. Lanning~386 could be
found at nearly any brightness between $V=17.4$ and 15.2  in 2005-2006. }
\label{allplot}
\end{figure}

\begin{figure}
\plotone{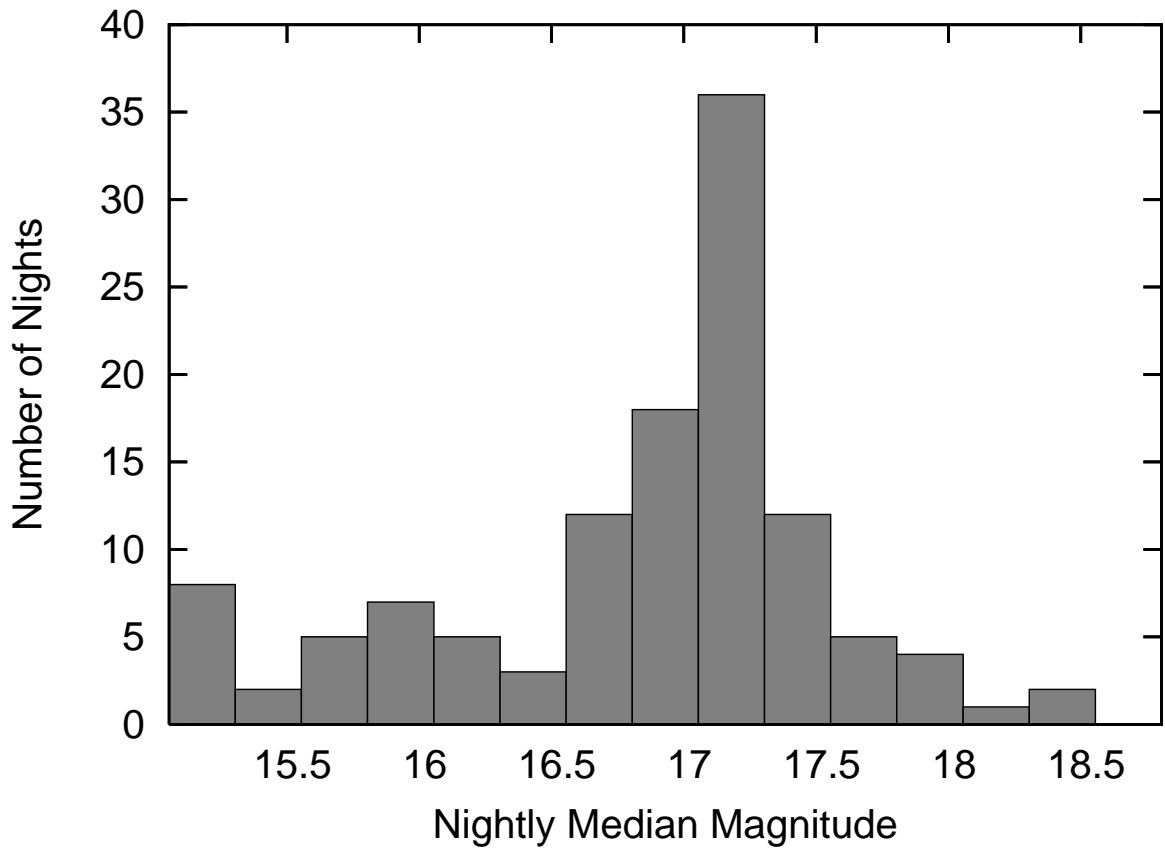}
\caption{Histogram of the median magnitude for each date with observations in 2005-2006. 
The star is most often found in a broad quiescent state between $V\approx 17.4$ and 16.7 mag,
but there are frequent bright outbursts to $V\approx 15.2$ mag.  Eclipse data are included,
and on a few nights the median magnitude corresponds to eclipse (see Figure \ref{detailplot}).
}
\label{histplot}
\end{figure}

\begin{figure}
\plotone{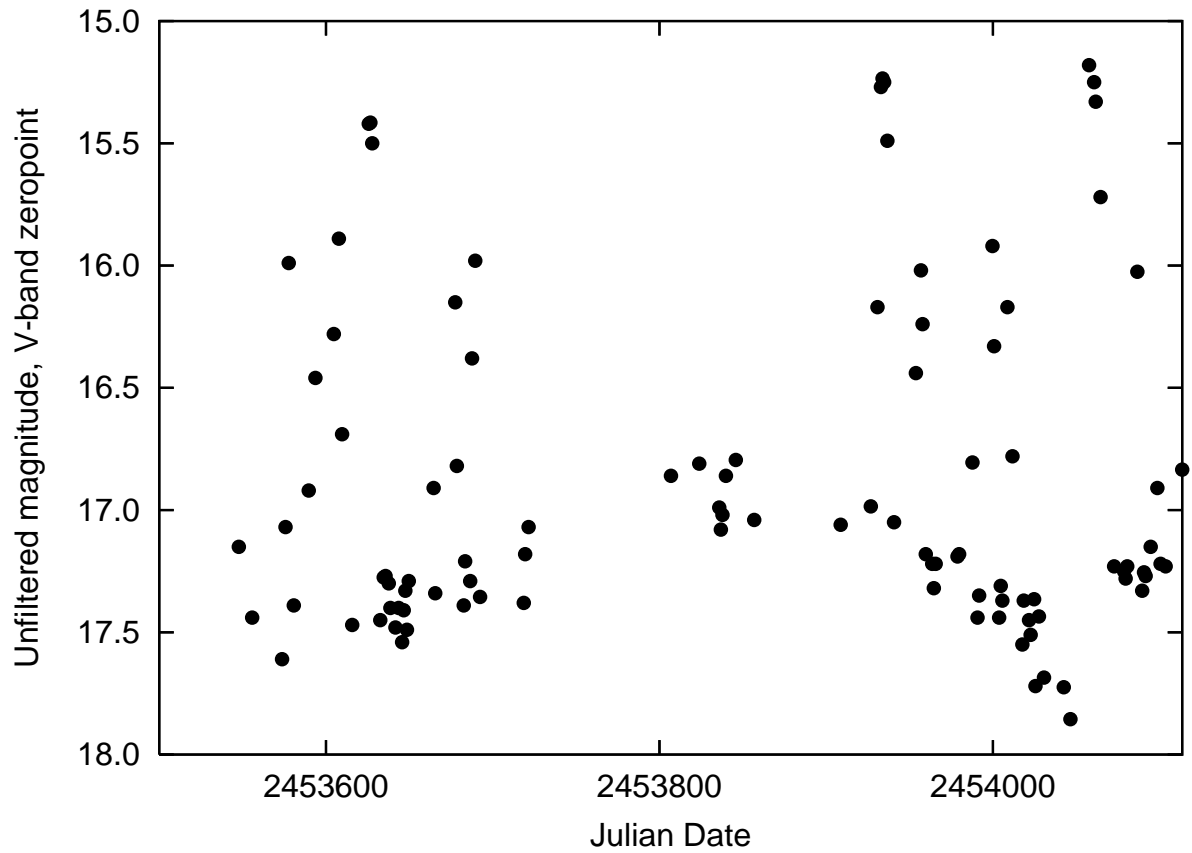}  % provisional -- from PDF.  Need a real postscript.
\caption{Nightly median magnitudes of Lanning 386, showing the irregular outbursts 
and intermediate states of Lanning~386.  The computation of the medians excluded
individual magnitudes taken in eclipse.}
\label{detailplot}
\end{figure}

Figure \ref{allplot} shows all the photometry folded on the ephemeris 
and makes evident the long term variations in the mean magnitude.
The histogram of the median magnitude for each of the dates in
Table \ref{photsumtable}, shown in Figure \ref{histplot}, suggests an
quiescent state near $m=17.2$ (with some variability), an intermediate stage at
$m=15.9$ and an outburst state at $m=15.3$.  This range of brightness
is also evident from visual inspection of Figure \ref{allplot}.
Figure \ref{detailplot} shows a typical portion of the long-term 
light curve,
from the 2006 observing season.  

%At first, this may appear to be simply a dwarf nova (DN) system, but
%Lanning~386 is very active spends much of its time in a bright or intermediate
%state. The light curve here is more similar to a very active VY~Scl object,
%for example MV~Lyr \citep{leach1999}.  These are bright nova-like stars that
%undergo fading that can last days to months. In particular, a number of SW~Sex
%type stars, such as DW~UMa \citep{honeycutt1993} show VY~Scl behavior.  The
%long-term light curve of LS~Peg (S193; Garnavich and Szkody 1992) for example,
%displayed periods of very rapid oscillations between bright and faint states
%that might be a sign of a transition from a low accretion rate state to a
%steady nova-like system.

\subsection{Spectral Appearance and Analysis}

\begin{figure} 
%\plotone{../decompspec.eps} 
\plotone{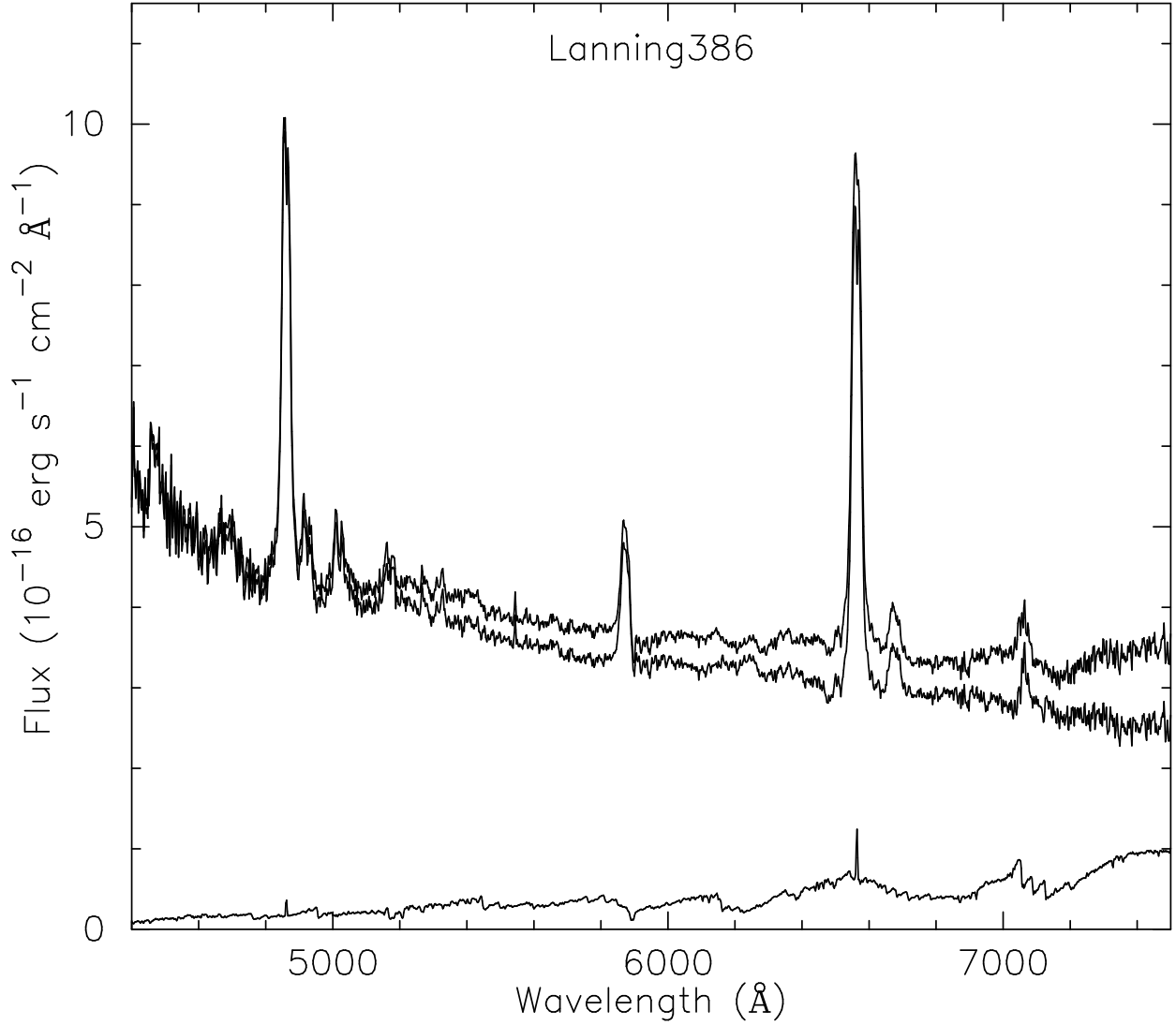}
\caption{{\it Top trace: } The mean spectrum of Lanning 386 in the low
state, from MDM Modspec observations. {\it Bottom trace:} A spectrum
of the M3.5e dwarf Gliese 388, scaled to 
$f_\lambda (6500\ {\rm \AA}) = 6 \times 10^{-17}$ 
erg cm$^{-2}$ s$^{-1}$ \AA $^{-1}$ to approximately
match the M-dwarf contribution in Lanning 386.  {\it Middle trace}
The difference of the top trace and the bottom trace.} 
\label{decompspec} 
\end{figure}

The top trace of Fig.~\ref{decompspec} shows the mean of the low-state
Modspec spectra.  The mean spectrum is typical of dwarf novae at
minimum light; Table \ref{emissionlines} gives identifications and
measurements of the emission features.  The synthetic magnitude is $V
= 17.5$, using the passband given by \citet{bessell}, which is 
roughly consistent with the low state photometry
outside of eclipse.  At the red end,
the absorption bands of an M dwarf are visible.  We estimated the
spectral class and flux contribution of the M dwarf by subtracting
scaled M-dwarf library spectra from the low-state spectrum and judging
how well the M-dwarf features were accounted for; this yielded a
spectral type M3.3 $\pm$ 0.8 and a contribution to the total light
corresponding to $V = 20.3 \pm 0.6$ (including estimated calibration
uncertainty).   Using (1) the measured $P_{\rm orb}$, (2)
the condition that the secondary fill the Roche lobe, and (3) a broad
range of plausible secondary star masses $M_2$ (from 0.2 to 0.5
M$_{\odot}$; \citealt{baraffe}), we estimate that the secondary's
scaled M-dwarf library spectra from the low-state spectrum and judging
how well the M-dwarf features were accounted for; this yielded a
spectral type M3.3 $\pm$ 0.8 and a contribution to the total light
corresponding to $V = 20.3 \pm 0.6$ (including estimated calibration
uncertainty).   Using (1) the measured $P_{\rm orb}$, (2)
the condition that the secondary fill the Roche lobe, and (3) a broad
range of plausible secondary star masses $M_2$ (from 0.2 to 0.5
M$_{\odot}$; \citealt{baraffe}), we estimate that the secondary's
radius $R_2 = 0.4 \pm 0.07$ R$_{\odot}$ -- fortunately, $R_2$ at a
fixed $P_{\rm orb}$ depends only weakly on $M_2$, roughly as
$M_2^{1/3}$.  Finally, using the calibration of the relation between
spectral type and $V$-band surface brightness given by
\citet{beuermann06}, we estimate $M_V = 11.1 \pm 1.0$ for the
secondary, giving an apparent distance modulus $m - M = 9.1 \pm 1.2$.
At Galactic latitude $b = -5.8$ there is likely to be substantial dust
extinction; \citet{schlegel} estimate that extragalactic objects in
this direction undergo reddening $E(B-V) = 0.54$.  The actual
reddening appears to be considerably less than this upper limit; most
dwarf novae in outburst have $B - V = 0$, within $\pm 0.1$ mag or so
\citep{bruch94}, so our outburst color suggests $E(B-V) < 0.2$ or so.
If so, the nominal distance would be $\sim 500$ pc.

\begin{figure}
\plotone{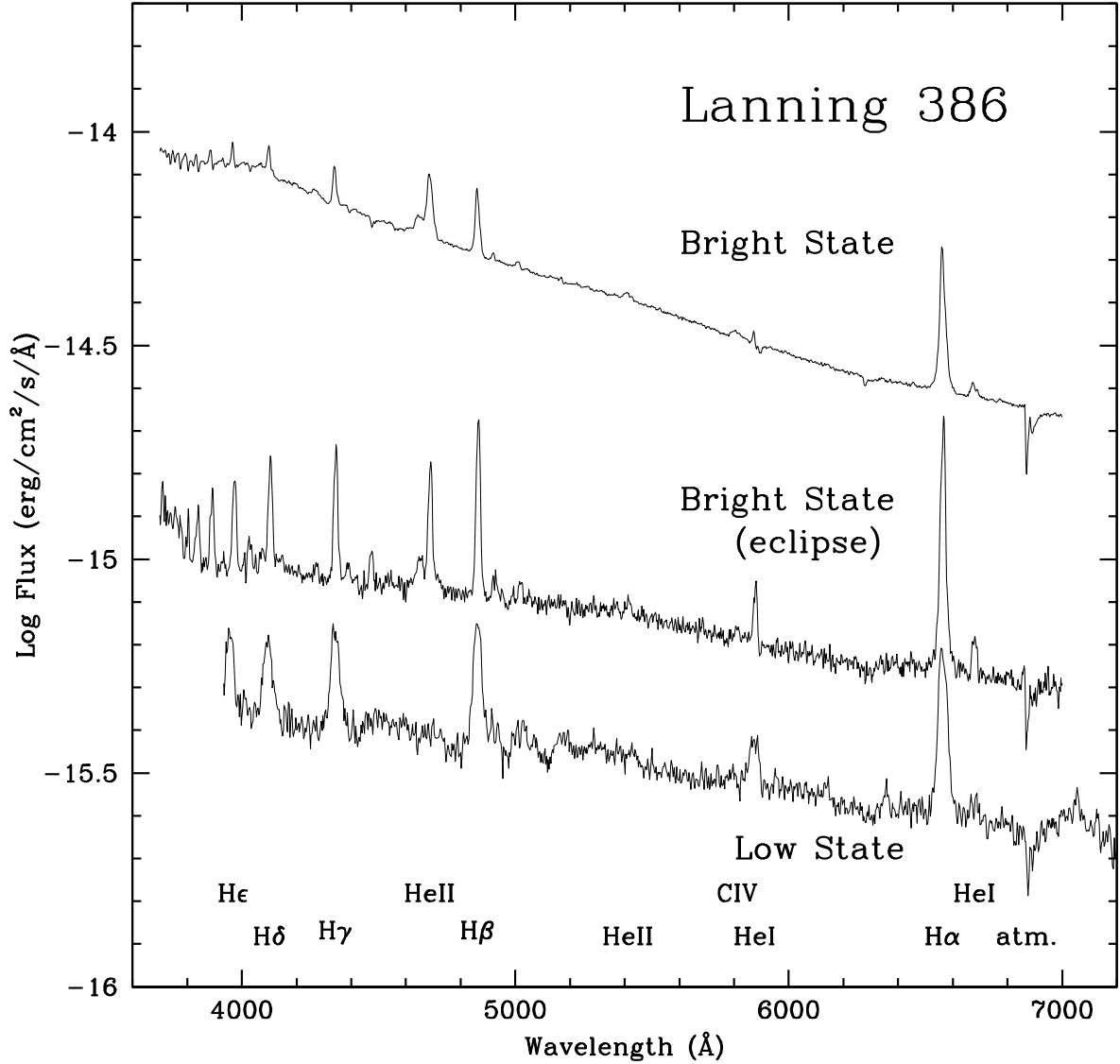}
\caption{{\it Top two traces:} The average of KPNO spectra taken outside eclipse
(top) plotted with the spectrum during eclipse (middle) and an MDM CCDS spectrum
taken at quiescence (bottom). The spectrum in the bright state shows a
very blue continuum outside of eclipse. In the bright state, Balmer
emission lines, emission lines of \ion{He}{1} and \ion{He}{2} and
broad emission from \ion{C}{4} are visible.  In quiescence, no high
excitation lines are detected. }
\label{spec1plot}
\end{figure}

The bright state KPNO spectra cover both eclipse and out-of-eclipse phases.
As shown in Figure \ref{spec1plot}, the average spectrum in
the bright state shows broad Balmer emission lines on a continuum that is
considerably bluer than in the low state.
Several \ion{He}{1} emission lines such as $\lambda\lambda 6678, 5876, 5016, 4922$  are
clearly seen, but the strongest non-hydrogen feature is the bright \ion{He}{2}
$\lambda4686$ line, which is blended with the \ion{C}{3}/\ion{N}{3} Bowen
complex. \ion{He}{2} is so prominent in this star that the usually weak line
at $\lambda5412$ is also easily detected.
A broad feature in the spectrum at $\lambda5806$ is evidently a
\ion{C}{4} feature which is common in Wolf-Rayet stars, but unusual in 
cataclysmic variables.
\citet{schmidtobreick2003} found the line in the old nova V840~Oph and
suggested the star had a carbon over-abundance. However, \citet{groot2000}
identified \ion{C}{4} in SW~Sex and attributed it to a particularly 
high-excitation state for the star.
In eclipse, the continuum remains blue, but its slope is greatly reduced.
Between 6000 \AA\  and 4000 \AA\ the continuum rises by a factor of 2.9 (in
$f_\lambda$) out of eclipse but only by a factor of 1.6 in eclipse.  However, in eclipse the
\ion{He}{2} emission remains one of the strongest lines in the spectrum.  The
ratio of fluxes between \ion{He}{2} $\lambda 4686$ and H$\beta$ is 1.2 out of eclipse
and 0.64 during eclipse.

If Lanning 386 is a dwarf nova, its high-state spectrum 
is unusual; we comment on this more extensively in the Discussion.

%The Balmer lines in dwarf novae typically become much weaker or go into absorption
%in outburst.  Even more strikingly, \ion{He}{2} emission is generally weak or
%absent, indicating a much lower excitation than seen here.  We discuss this
%points more thoroughly later.

%In quiescence, Balmer and \ion{He}{1} emission lines are prominent. But in
%this low state the continuum slope matches that of the star in eclipse and
%there is no detectable \ion{He}{2} emission. The equivalent width (EW) of the
%Balmer emission lines shows large variations in and out of eclipse and during
%quiescence.  The average EW of H$\beta$ outside of eclipse in the bright 
%state is 8~\AA\footnote[1]{Here we are using positive equivalent widths to denote
%emission lines.} while at mid-eclipse the EW reaches 34~\AA , comparable
%to the faint state.

%The average KPNO bright-state spectrum out of eclipse (phases 0.4 to nearly 1.0) shows
%\ion{He}{1} $\lambda$4471 in absorption, while the redward \ion{He}{1} lines are in
%emission. 
%The low resolution of the spectra make it difficult to check for the
%zero-velocity absorption on the Balmer lines at phase 0.5 which is
%characteristic of SW~Sex stars. 
%But the \ion{He}{1} $\lambda$4471 absorption moves from blue to red 
%between phase 0.4 to 0.6 and,
%\ion{He}{1} $\lambda 6678$ has a central absorption over this phase range.

\begin{figure}
% \epsscale{0.70}
\epsscale{0.68}
%\plotone{singletrail.eps}
\plotone{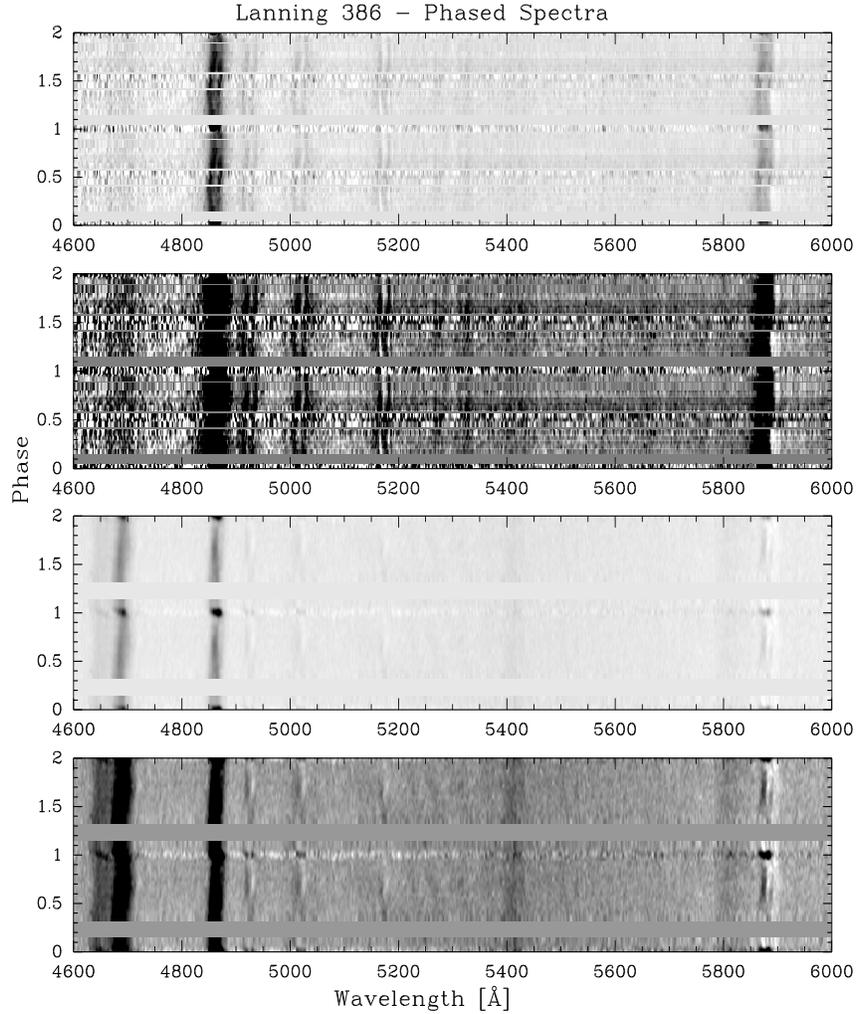}
\caption{
Phase-averaged greyscale representations of the MDM Modspec low-state
spectra (top two panels) and the KPNO high-state spectra (lower two panels).  
In the top of each image pair the greyscale is selected to make the details of 
the strong lines visible, while the lower image is scaled to show the fainter
features.  The procedure used
to create this is described in the text.  Blank horizontal bars appear 
where there are no spectra close enough in phase to include, and other
horizontal features are mostly from inadequate phase coverage, except for the
eclipse, which is the low signal-to-noise stripe at phase 1.  The
apparent increase of the emission strength through eclipse is an artifact of the
normalization used.
}
\label{singletrail}
\end{figure}

Figure~\ref{singletrail} shows phase-resolved greyscale representations of our
high- and low-state spectra.  To construct these, the spectra were first divided
by the continuum and cleaned of residual cosmic rays; then for each orbital
phase we averaged spectra whose phases were close to the target phase, using
a Gaussian in phase for the weighting function.  Finally, all the spectra
were stacked together to form a two-dimensional image, with a second cycle 
repeated for continuity.  We created separate images from the high-state
KPNO data and the low-state MDM Modspec data.

Examination of Fig.~\ref{singletrail} shows a host of interesting differences
between the two states.   In the low state (top two panels), the emission lines
have two approximately equal peaks, while in the high state (lower two panels)
show more complex behavior; in particular, phase-dependent absorption is seen in
\ion{He}{1} $\lambda\lambda$4921, 5015, and 5876, and in the metal feature
near $\lambda 5175$.  In H$\beta$ (and in H$\alpha$, which is not shown here)
there is a sharp component that moves approximately in antiphase to the bulk
of the line.  The phasing of the sharp component's motion is approximately
as expected for emission from the heated face of the 
secondary.  The high-state image has good coverage across the eclipse at
phase $\phi = 0$; because the emission lines are eclipsed less strongly than the
continuum, the continuum division used in this representation causes an 
increase in the apparent emission strength.  This is artificial, but the 
sharp red-to-blue swing in the line's central wavelength is evidently 
real (see below).

% Figure \ref{spec2plot} we average four spectra taken near phase 0.75
%showing inverse P-Cygni absorption for all the \ion{He}{1} lines and for
%Balmer lines starting with H$\delta$. Complex interaction between emission and
%absorption components like this are characteristic of SW~Sex stars
%\citep{thorstensen1991}. In particular, the inverse P-Cygni profile of 
%\ion{He}{1} $\lambda$4471 is similar to V795~Her \citep{casares96},
%although the \ion{He}{2} emission in Lanning~386 is stronger than in V795~Her.
%The unusual variable absorption feature at $\lambda$5175 noted in the SW~Sex star
%PX And (=PG0027+260; \citealt{thorstensen1991}) and identified as \ion{Mg}{1} is also
%present in Lanning~386.

\section{Radial Velocities} 

We measured radial velocities of the H$\alpha$ emission in the
Modspec and KPNO spectra using a convolution algorithm described by
\citet{schyo} and \citet{shaf}.  For the low-state Modspec data the
convolving function consisted
of positive and negative Gaussians with a full-width at half
maximum (FWHM) of 550 km s$^{-1}$, separated
by 1920 km s$^{-1}$, which emphasized the steep sides of the line profile. 
The corresponding figures for the slightly narrower line profile
in the high-state data were respectively 365 and 1645 km s$^{-1}$. 

\begin{figure}
\epsscale{1.0}
%\plotone{../velfolpl.eps}
\plotone{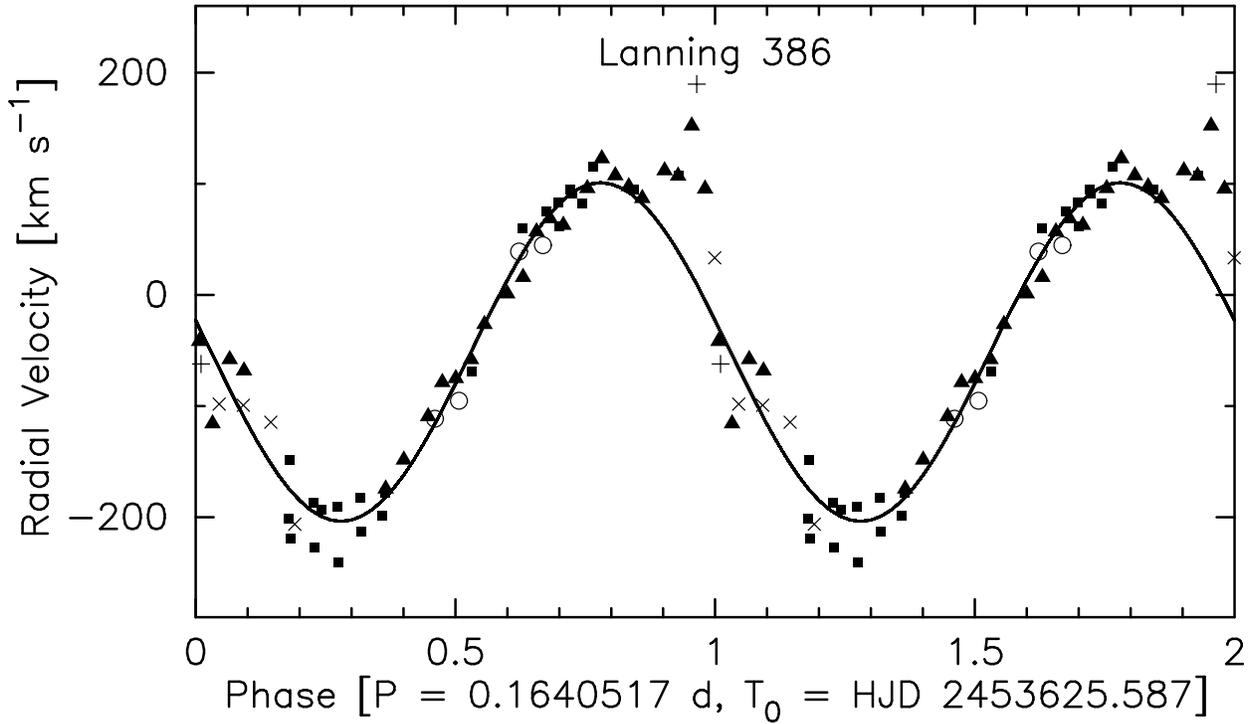}
\caption{Radial velocities of H$\alpha$ from the MDM Modpsec and KPNO
4m, folded using the eclipse ephemeris.  The solid triangles are the
KPNO high-state data; the X-crosses are from 2005 Sep high-state
spectra; the remainder were all taken in the low state, and the
symbols denote different observing runs as follows: plus-signs = 2005
July; solid squares = 2006 Aug/Sep; open circles = 2007 Mar/Apr.  All
the velocities are re-plotted for a second cycle to preserve
continuity.  Note the prominent rotational disturbance around the
eclipse phase.
}
\label{folvel}
\end{figure}

We searched the velocity time series for periods; this exercise 
independently reproduced the same
period found in the eclipses, with no cycle-count ambiguity on any
time scale.  Remarkably, the high- and low-state velocities
agreed very well in phase, amplitude, and zero point, despite
differences in instrumentation as well as photometric state.
The modulation is accurately sinusoidal, {\it except} for 
a rapid swing redward and then blueward across the eclipse --
the classic signature of the eclipse of a rapidly-rotating
object.  Excluding the velocities affected by the rotational
disturbance, holding $P$ fixed at $0.1640517$ d,
and fitting a sinusoid of the form
$v(t) = \gamma + K \sin [2 \pi (t - T_0)/ P]$
yields
\begin{eqnarray}
T_0 &=& \hbox{HJD } 2453980.6815 \pm 0.0010, \\
K &=& 152 \pm 5\ {\rm km\ s^{-1}}, \\
\gamma &=& -51 \pm 4\ {\rm km\ s^{-1}}, \hbox{ and}\\
\sigma &=& 18\ {\rm km\ s^{-1}},
\end{eqnarray}
where $\sigma$ is the RMS scatter about the best fit. 
Fig.~\ref{folvel} shows the velocities folded on the eclipse
ephemeris.

The sinusoid fitted to the velocities is defined so that 
$T_0$ is the time at which $v(t)$ crosses the system
velocity $\gamma$ from blue to red.  If the emission lines trace the
motion of the white dwarf, this must occur opposite
the eclipse, at phase $\phi = 0.5$.  The fitted
$T_0$ corresponds to $\phi = 0.530 \pm 0.006$, or a phase lag of
$11^{\circ} \pm 2^{\circ}$ compared to this expectation.  This suggests that the
emission line velocities do not trace the white dwarf motion precisely.

\section{Discussion}

Our photometry shows Lanning~386 to be an active, eclipsing CV, with $P_{\rm orb} = 
3.94$ h and an eclipse depth of $\sim 1.5$ mag. 
Monitoring over many months shows bright outbursts reaching $V\approx 15.2$
from a quiescent state with $V\approx 17.2$. 
%But the star has been caught
%between this range and may spend some time in a intermediate state around
%$V\approx$15.5 mag. 
%The behavior of the light curve is not like that of dwarf
%novae, but appears to similar to rapid oscillations seen in VY~Scl and some
%SW~Sex stars like LS~Peg.

Lanning 386 does not fit neatly into any CV subclass.
The long-term variability resembles that of a dwarf nova, and at 
minimum light the spectrum is typical of that class.
However, when most dwarf novae outburst, their Balmer emission
becomes very weak, and broad absorption wings appear
(e.g., \citealt{thorstensen98}); 
it is also unusual (though not unknown) for \ion{He}{2} to 
become prominent (e.g., \citealt{hessman84}).
The high-state spectrum seen here is very much unlike this,
resembling instead that of a novalike variable.
The \ion{He}{1} line profiles, and the unusual strength of 
\ion{He}{2}, are reminiscent of an SW Sextantis star \citep{thorstensen1991}.
In addition, Lanning 386 is similar to other SW Sex stars in that 
the emission lines are eclipsed much more shallowly than the continuum,
indicating a spatially extended source.  The 
rotational disturbance around eclipse is also seen in some
other SW Sex stars (e.g., \citealt{rodriguez04}).
However, other canonical SW Sex characteristics are missing here.
The emission line radial-velocity phase in SW Sex stars lags
far behind (by $> 45^{\circ}$) the phase expected for the white
dwarf motion, based on the eclipse ephemeris, but in Lanning 386 
the phase lag is only $\sim 11^\circ \pm 2^\circ$.  
In addition, while the \ion{He}{1} lines do appear in absorption
for part of the orbit, in other SW Sex stars the absorption appears
near phase 0.5, whereas it appears somewhat later here.

Nonetheless, the unusual maximum-light spectrum suggests that we should
consider carefully whether the long-term light curve in Fig.~\ref{detailplot}
forces us to classify this star as a dwarf nova.  Might another mechanism
explain the variability in a manner more consistent with the maximum-light
spectrum?  

Many novalike variables, and many SW Sex stars in particular, are
also VY Scl stars -- novalike variables that occasionally fade deeply for
varying amounts of time, for reasons that are not entirely understood.  Might
Lanning 386 be a VY Scl star masquerading as a dwarf nova?   This seems
unlikely, for two reasons.  First, the long-term light curves of VY Scl stars
tend to look quite different from those of dwarf novae -- VY Scl stars
typically remain in a high state for months or years, and only occasionally
fade by large amounts, up to 5 mag (see examples in \citealt{honeycutt04}).
The light curve here, by contrast, consists of frequent, moderate-amplitude
outbursts from a faint state.  Second, the spectra of VY Scl stars in their
faint states look as if accretion has nearly stopped, with the Balmer emission
lines typically becoming very narrow and the underlying stars becoming evident
\citep{shafter1983, schneider81}.  This is very different from the essentially
normal dwarf nova spectrum seen here.  

The DQ Her stars, also called intermediate polars, are 
magnetic systems \citep[for a review, see][]{patterson94}; some of these
undergo outbursts that at least superficially resemble dwarf nova
outbursts.
Typically, DQ Her outbursts are infrequent compared to those of most
dwarf novae; they also tend to decay rapidly \citep{hellier97}.  
For this reason, there is little spectroscopy available for outbursts
of these systems.  However, spectra do exist for EX Hya in outburst, which,
show strong emission lines with interesting 
velocity structure, reminiscent of the SW Sex stars 
\citep{hellier89,hellier00}.  In this way, EX Hya resembles
Lanning 386, but the morphological fit is 
imperfect -- the brightenings we see in Lanning 386 appear
to be more frequent than those seen in most DQ Her stars.
Also, we have (as of yet) no direct evidence that Lanning 386 is 
a magnetic system, because we are unaware of any 
searches for the coherent pulsations and/or 
circular polarization that would prove the case.  

It has been suggested that the outbursts seen in some magnetic systems are
due to mass-transfer bursts rather than the disk instabilities that are
thought to cause normal dwarf nova outbursts 
\footnote{It is possible that mass-transfer bursts could be triggered
by an event in the disk.}.  During a period of increased mass transfer,
material overflowing the disk could give rise to enhanced emission --
there is evidence for this in EX Hya \citep{hellier00}.  
It may be that the outbursts seen in Lanning 386, whatever their
trigger, involve enhanced mass transfer -- which may or may not
be exclusive to magnetic systems.  

We can summarize the morphological conundrum presented by Lanning 386 as
follows:  Its minimum-light spectrum and the light curve suggest it
is a dwarf nova with frequent, relatively low-amplitude outbursts, 
while at maximum light it resembles a novalike variable, with some 
(but not all) the characteristics of an SW Sex star.  EX Hya, 
a magnetic system, shows a similar set of symptoms, but the
presentation is not identical.

\section{Acknowledgments}

While this paper was in production we learned that H. Lanning had died
in 2007 December; we thank him posthumously 
for encouraging us to get spectra of this star and dedicate this
paper to his memory.  JRT 
acknowledges support from the National Science Foundation through 
awards AST-0307413 and AST-0708810.  We thank the KPNO and MDM staffs
for their conscientious support, and the Tohono O'odham Nation for
letting us use their mountain for a while to explore the universe we
all share.  Finally, we thank the referee for a careful 
reading and numerous helpful suggestions.

%
% Begin tables
%

\begin{deluxetable}{lcl}
\tabletypesize{\normalsize}
\tablecaption{Known or Suspected Lanning CVs\label{lanningtable}}
\tablecolumns{13}
\tablewidth{0pt}
\tablehead{
\colhead{Object} & \colhead{Classification} & \colhead{Source}\\
}
\startdata
Lanning 10 & nl/ux & \citet{Horne1982}\\
Lanning 17 & n: & \citet{lanning1973}\\
Lanning 90 & ux & \citet{shafter1983}\\
Lanning 159\tablenotemark{a} & nl: & \citet{eracleous2002}\\
Lanning 302 & n: & \citet{lanning1998}\\
Lanning 386 & CV: & \citet{eracleous2002}\\
Lanning 420 & n: & \citet{lanning2000}\\
\enddata
\tablenotetext{a}{\citet{liu00} obtained a spectrum of Lanning 
159 that did not support a CV classification; one of us (JRT) 
also obtained a
spectrum with the MDM 2.4 m, using the same MDM modspec
setup described in this paper, which at low signal-to-noise
shows H$\alpha$ in absorption with a profile resembling 
a DA white dwarf.}
\end{deluxetable}	

% REVISED table from Steve -- 2007 Aug 22

\begin{deluxetable}{crrc}
\tabletypesize{\normalsize}
\tablecaption{Comparison Stars for Lanning 386\label{cmptable}}
\tablecolumns{4}
\tablewidth{0pt}
\tablehead{
\colhead{Star} & 
\colhead{$\alpha_{2000}$\tablenotemark{a}} & 
\colhead{$\delta_{2000}$} & 
\colhead{V} \\
\colhead{} & \colhead{[hh:mm:ss]} & \colhead{[$^{\circ}$:$'$:$''$]} & \colhead{} \\
}
\startdata
A & 21:08:29.82 & 39:04:43.7 & 14.633 \\
B & 21:08:31.94 & 39:04:47.1 & 15.759 \\
C & 21:08:32.53 & 39:05:44.0 & 16.983 \\
D & 21:08:40.15 & 39:05:06.6 & 15.976 \\
\enddata
\tablenotetext{a}{Celestial coordinates are measured from MDM 2.4m images,
using a coordinate solution based on numerous UCAC2 stars \citep{zacharias04}.}
\end{deluxetable}

\begin{deluxetable}{lcccl}
\tabletypesize{\tiny}
\tablecaption{Log of Photometric Observations\label{photsumtable}}
\tablecolumns{13}
\tablewidth{0pt}
\tablehead{
\colhead{UT Date} & \colhead{Duration(hr)} & \colhead{Avg mag outside eclipse} & \colhead{Eclipse depth} & \colhead{Comments}\\
}
\startdata
% Data have been edited to add 0.4 to each magnitude -- JRT 2007 Aug 30
2005 Sep 12	&	5.8	&      15.4    &	1.4	&	Large amplitude outburst	\\
 
2005 Sep 13	&	5.2	&      15.4    &	1.4	&	Large amplitude outburst	\\
 
2005 Sep 14	&	6.6	&      15.5    &	1.5	&	Large amplitude outburst, Small hump??	\\
 
2005 Sep 22	&	6.6	&      17.3    &	1.2	&	Quiet, Strong 0.3 mag hump	\\
 
2005 Sep 24	&	6.0	&      17.3    &	1.4	&	Quiet, Strong 0.4 mag hump	\\
 
2005 Sep 25	&	6.6	&      17.4    &	1.2	&	Quiet, Strong 0.4 mag hump	\\
 
2005 Sep 28	&	6.5	&      17.4    &	1.5	&	Quiet, Small 0.3 mag hump	\\
 
2005 Oct 02	&	6.4	&      17.5    &	1.3	&	Quiet, Strong 0.4 mag hump	\\
 
2005 Nov 03	&	3.9	&      16.1    &	2.0	&	Minor outburst	\\
 
2005 Nov 08	&	3.7	&      17.4    &	1.6	&	Quiet	\\
 
2005 Nov 13	&	4.1	&      16.3    &	1.6	&	Minor outburst, 0.3 mag hump	\\
 
2006 Jul 10	&	5.8	&      16.9    &	2.0	&	Quiet	\\
 
2006 Jul 17	&	6.0	&      15.2    &	1.4	&	Major outburst	\\
 
2006 Jul 18	&	5.8	&      15.2    &	1.6	&	Major outburst	\\
 
2006 Jul 20	&	6.4	&      15.5    &	1.7	&	Major outburst, fading	\\
 
2006 Jul 24	&	5.7	&      17.0    &	2.1	&	Quiet	\\
 
2006 Aug 09	&	4.6	&      16.0    &	2.3	&	Minor outburst	\\
 
2006 Aug 10	&	7.4	&      16.2    &	2.0	&	Minor outburst	\\
 
2006 Aug 31	&	2.2	&      17.2    &	1.8	&	Quiet	\\
 
2006 Sep 01	&	5.8	&      17.2    &	1.2	&	Quiet, 0.4 mag hump	\\
 
2006 Sep 09	&	6.9	&      16.8    &	1.6	&	Minor flare up or fading	\\
 
2006 Sep 12	&	7.3	&      17.4    &	1.5	&	Quiet, 0.3 mag hump	\\
 
2006 Sep 13	&	6.7	&      17.3    &	1.5	&	Quiet, 0.3 mag hump	\\
 
2006 Sep 21	&	6.4	&      15.9    &	2.0	&	Minor outburst	\\
 
2006 Sep 22	&	6.4	&      16.3    &	1.9	&	Minor outburst, fading	\\
 
2006 Sep 25	&	6.7	&      17.5    &	1.3	&	Quiet, Strong 0.5 mag hump	\\
 
2006 Sep 27	&	4.3	&      17.4    &	1.4	&	Quiet, 0.3 mag hump	\\
 
2006 Sep 30	&	6.0	&      16.2    &	2.1	&	Minor outburst	\\
 
2006 Oct 03	&	6.6	&      16.8    &	1.9	&	Minor outburst/ fading	\\
 
2006 Oct 09	&	5.8	&      17.5    &	1.4	&	Quiet, Strong 0.4 mag hump	\\
 
2006 Oct 13	&	3.9	&      17.5    &	1.4	&	Quiet, 0.3 mag hump	\\
 
2006 Oct 22	&	6.0	&      17.7    &	1.0	&	Quiet, 0.3 mag hump	\\
 
2006 Nov 18	&	3.5	&      15.1    &	2.2	&	Large amplitude outburst	\\
 
2006 Nov 21	&	4.8	&      15.2    &	1.5	&	Large amplitude outburst	\\
 
2006 Nov 24	&	5.2	&      15.6    &	1.8	&	Large amplitude outburst, fading	\\
 
2006 Dec 03	&	4.3	&      17.2    &	1.4	&	Quiet, 0.3 mag hump	\\
 
2006 Dec 08	&	3.4	&      17.2    &	1.5	&	Quiet	\\
 
2006 Dec 09	&	3.0	&      17.2    &	1.5	&	Quiet	\\
 
2006 Dec 10	&	4.1	&      17.2    &	1.5	&	Quiet, 0.3 mag hump	\\
 
2006 Dec 11	&	3.1	&      17.2    &	1.5	&	Quiet, 0.3 mag hump	\\
 
2006 Dec 17	&	2.4	&      16.0    &	1.8	&	Minor outburst	\\
 
2006 Dec 19	&	2.6	&      17.3    &	1.4	&	Quiet	\\
 
2006 Dec 20	&	2.0	&      17.2    &	1.4	&	Quiet, 0.2 mag hump	\\
 
2006 Dec 21	&	2.0	&      17.2    &	1.4	&	Quiet	\\
 
2006 Dec 25	&	2.8	&      17.1    &	1.6	&	Quiet
 
\enddata
\end{deluxetable}

%\begin{deluxetable}{rrr}
%\tabletypesize{\normalsize}
%\tablecaption{Dominant frequencies from Fourier analysis.\label{freqs}}
%\tablecolumns{13}
%\tablewidth{0pt}
%\tablehead{
%\colhead{Frequency} & \colhead{Period} & \colhead{Amplitude}\\
%}
%\startdata
%$6.09537 \pm 0.00016$ & $0.164059 \pm 0.000004$ & $0.247 \pm 0.041$\\
%$12.19128 \pm 0.00016$ & $0.082026 \pm 0.000001$ & $0.254 \pm 0.041$\\
%$3.08371 \pm 0.00014$ & $0.324284 \pm 0.000015$ & $0.285 \pm 0.041$\\
%\enddata
%\end{deluxetable}

\begin{deluxetable}{lrcccr}
\tabletypesize{\scriptsize}
\tablewidth{0pt}
\tablecolumns{6}
\tablecaption{Journal of Spectroscopic Observations\label{specjournal}}
\tablehead{
\colhead{Date} &
\colhead{$N$} &
\colhead{HA start} &
\colhead{HA end} &
\colhead{Instrument\tablenotemark{a}} & 
\colhead{state} \\
\colhead{(UT)} &
\colhead{ } &
\colhead{[hh:mm]} &
\colhead{[hh:mm]} &
\colhead{ } &
\colhead{ } \\
}
\startdata
2005 Jul 01  &  2  &  $-$0:15  &  $-$0:04 & mod & low\\
2005 Sep 09  &  5  &    +1:42  &    +2:27 & mod & high\\
2006 Aug 31  &  5  &  $-$0:17  &    +4:35 & mod & low\\
2006 Sep 01  &  11 &  $-$1:15  &    +5:58 & mod & low\\
2006 Sep 02  &  7  &  $-$2:57  &  $-$0:38 & mod & low\\
2006 Nov 18  &  25 &  $+$1:28  &  $+$4:20 & RC  & high\\
2006 Dec 08  &  2  &   +1:55   &   +2:07  & CCDS & low \\  % HA is faked sligktly.
2007 Mar 30  &  2  &  $-$4:01  &  $-$3:50 & mod & low\\
2007 Apr 01  &  2  &  $-$4:00  &  $-$3:49 & mod & low\\
\enddata
\tablenotetext{a}{mod = MDM 2.4m Hiltner + modspec; RC = KPNO 4m Hiltner + 
RC spectrograph; CCDS = MDM 2.4m Hiltner + CCDS}
\label{tab:obsjournal}
\end{deluxetable}

\begin{deluxetable}{lccl}
\tabletypesize{\tiny}
\tablecaption{Observed Eclipses of Lanning 386\label{eclipsestable}}
\tablecolumns{13}
\tablewidth{0pt}
\tablehead{
\colhead{HJD(+2,450,000)} & \colhead{Epoch} & \colhead{O-C (sec)} & \colhead{UT Date}\\
}
\startdata
3625.58669	$\pm$	0.00044	&	0		&	0  	& 2005 Sep 12 \\
3625.75074	$\pm$	0.00091	&	1	 	&	27		&	2005 Sep 12 \\
3626.73505	$\pm$	0.00041	&	7		&	59		&	2005 Sep 13 \\
3627.55531	$\pm$	0.00002	&	12		&	53		&	2005 Sep 14 \\
3627.71936	$\pm$	0.00044	&	13	 	&	40		&	2005 Sep 14 \\
3635.59385	$\pm$	0.00049	&	61	 	&	42		&	2005 Sep 22 \\
3635.75790	$\pm$	0.00156	&  62	 	&	31		&	2005 Sep 22 \\
3637.56247	$\pm$	0.00002	&	73	 	&	135	&	2005 Sep 24 \\
3637.72652	$\pm$	0.00002	&	74	 	&	77		&	2005 Sep 25 \\
3641.66376	$\pm$	0.00067	&	98	 	&	80		&	2005 Sep 28 \\
3645.60100	$\pm$	0.00002	&	122 	&	15		&	2005 Oct 02 \\
3677.59108	$\pm$	0.00058	&	317 	&	39		&	2005 Nov 03 \\
3682.51263	$\pm$	0.00002	&	347 	&	77		&	2005 Nov 08 \\
3687.59823	$\pm$	0.00002	&	378   &	63		&	2005 Nov 13 \\
3933.67578	$\pm$	0.00002	&	1878  &	53		&	2006 Jul 17 \\
3934.66010	$\pm$	0.00002	&	1884  &	53		&	2006 Jul 18 \\
3934.82415	$\pm$	0.00021	&	1885	&	22		&	2006 Jul 18 \\
3936.62872	$\pm$	0.00002	&	1896	&	27		&	2006 Jul 20 \\
3936.79277	$\pm$	0.00002	&	1897	&	21		&	2006 Jul 20 \\
3940.73001	$\pm$	0.00352	&	1921	&	104 	&	2006 Jul 24 \\
3956.80707	$\pm$	0.00159	&	2019	&	44  	&	2006 Aug 09 \\
3957.62733	$\pm$	0.00098	&	2024	&	20		&	2006 Aug 10 \\
3957.79138	$\pm$	0.00056	&	2025	&	26		&	2006 Aug 10 \\
3979.77431	$\pm$	0.00067	&	2159	&	22		&	2006 Sep 01 \\
3987.64879	$\pm$	0.0012	&	2207	&	-30	& 2006 Sep 09 \\
3990.60172	$\pm$	0.00002 	&	2225	&	20		&	2006 Sep 12 \\
3991.58603	$\pm$	0.00517	&	2231	&	-30	&	2006 Sep 113\\
3991.75009	$\pm$	0.00076	&	2232	&	39		&	2006 Sep 13 \\
3999.62457	$\pm$	0.00048	&	2780	&	302	&	2006 Sep 21 \\
4000.60888	$\pm$	0.00145	&	2286	&	18		&	2006 Sep 22 \\
4003.56181	$\pm$	0.0016	&	2304	&	36		&	2006 Sep 25 \\
4003.72586	$\pm$	0.00145	&	2305	&	40		&	2006 Sep 25 \\
4005.53043	$\pm$	0.00074	&	2316	&	79		&	2006 Sep 27\\
4005.69448	$\pm$	0.00123	&	2317	&	53		&	2006 Sep 27\\
4008.64741	$\pm$	0.00076	&	2335	&	36		&	2006 Sep 30\\
4011.60034	$\pm$	0.00087	&	2353	&	35		&	2006 Oct 03\\
4017.50620	$\pm$	0.00058	&	2389	&	46		&	2006 Oct 09\\
4017.67026	$\pm$	0.00102	&	2390	&	44		&	2006 Oct 09\\
4022.59181	$\pm$	0.0009	&	2420	&	48		&	2006 Oct 13\\
4030.63034	$\pm$	0.0018	&	2469	&	-37	&	2006 Oct 22\\
4057.53482	$\pm$	0.00156	&	2633	&	-3		&	2006 Nov 18\\
4064.58904	$\pm$	0.00035	&	2676	&	30		&	2006 Nov 24\\
4072.46352	$\pm$	0.00083	&	2724	&	72		&	2006 Dec 03\\
4078.53344	$\pm$	0.0012  	&	2761	&	60		&	2006 Dec 09\\
4079.51775	$\pm$	0.00096	&	2767	&	29		&	2006 Dec 10\\
4089.52490	$\pm$	0.002	  	&	2828  &	64		&	2006 Dec 20\\
4090.50921	$\pm$	0.00107	&	2834	&	10		&	2006 Dec 21\\
4091.49352	$\pm$	0.00091	&	2840	&	115	&	2006 Dec 21\\
4094.44645	$\pm$	0.00077	&	2858	&	60		&	2006 Dec 24\\
\enddata
\end{deluxetable}

\begin{deluxetable}{lrcc}
\tabletypesize{\scriptsize}
\tablewidth{0pt}
\tablecolumns{4}
\tablecaption{Emission Lines in Quiescence}
\tablehead{
&
\colhead{E.W.\tablenotemark{a}} &
\colhead{Flux}  &
\colhead{FWHM \tablenotemark{b}} \\
\colhead{Feature} &
\colhead{(\AA )} &
\colhead{(10$^{-16}$ erg cm$^{-2}$ s$^{1}$)} &
\colhead{(\AA)} \\
}
\startdata

           H$\gamma$ & $ 27$ & $166$ & 28 \\ 
  HeI $\lambda 4471$ & $  6$ & $ 31$ & 36 \\ 
            H$\beta$ & $ 46$ & $196$ & 30 \\ 
  HeI $\lambda 4921$\tablenotemark{c} & $  8$ & $ 32$ & 36 \\ 
  HeI $\lambda 5015$\tablenotemark{c} & $  7$ & $ 32$ & 46 \\ 
  FeII$\lambda 5169$ & $  4$ & $ 17$ & 38 \\ 
  HeI $\lambda 5876$ & $ 10$ & $ 39$ & 26 \\ 
           H$\alpha$ & $ 65$ & $227$ & 32 \\ 
  HeI $\lambda 6678$ & $  6$ & $ 20$ & 33 \\ 
  HeI $\lambda 7067$ & $  6$ & $ 20$ & 37 \\ 
\label{emissionlines}
\enddata

\tablenotetext{a}{Emission equivalent widths are counted as positive.}
\tablenotetext{b}{From Gaussian fits.}
\tablenotetext{c}{The $\lambda 4921$ and 5015 features are generally
ascribed to HeI, but may include a contribution from \ion{Fe}{2} 
$\lambda\lambda 4924$ and 5018.}

\end{deluxetable}

\clearpage

%
% Begin figures
%

%\begin{figure} 
%\plotone{f1.eps} 
%\caption{Finder chart for Lanning 386 taken near maximum light with 
%the 0.25 m survey telescope.  North is up and east to the left, and 
%the field of view is 5 arcmin square.}
%\label{finder_close} \end{figure}

\begin{figure}
\plotone{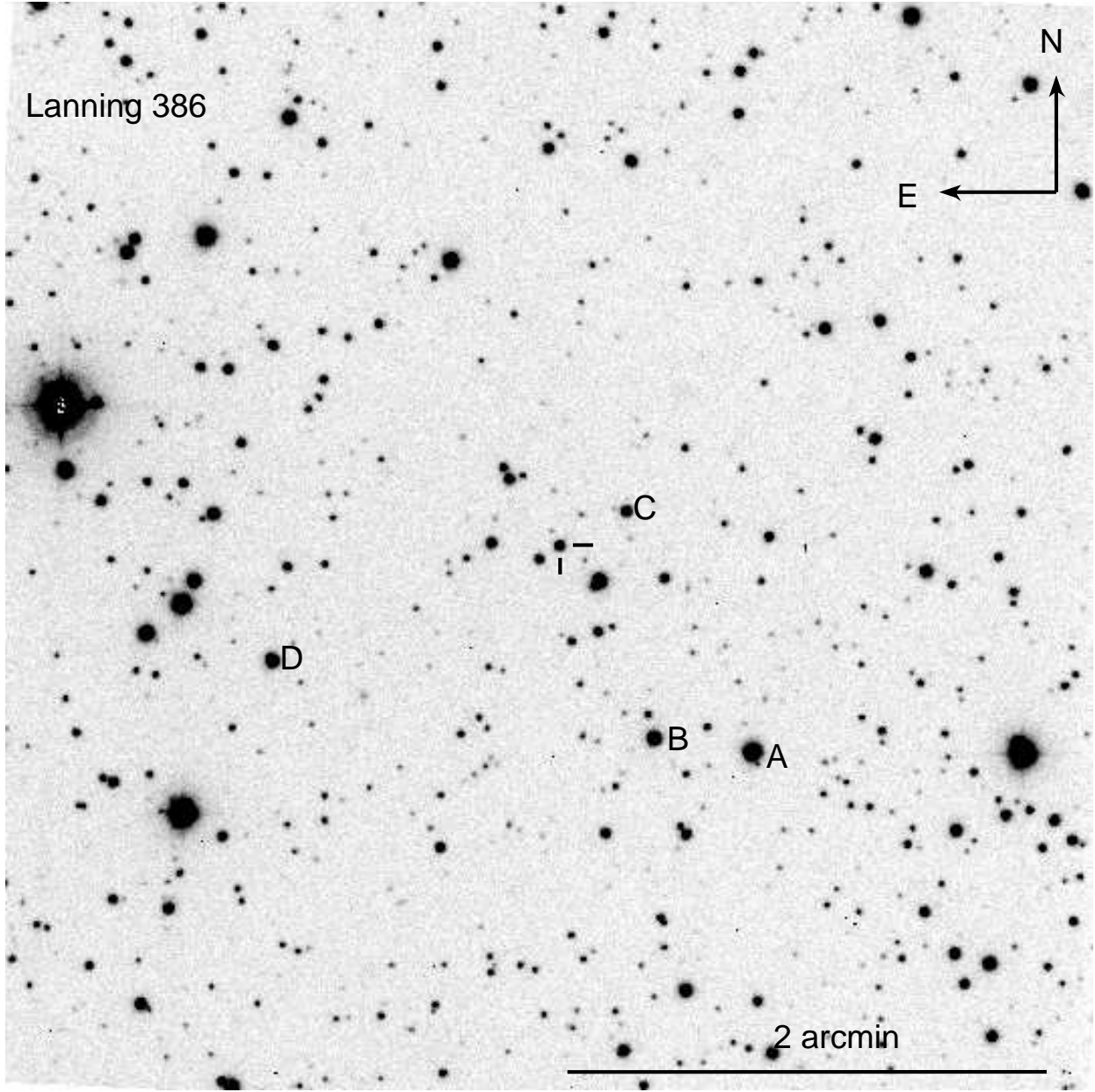}
\caption{
A 60-second $V$-band image of Lanning 386 taken with the 2.4m MDM Hiltner telescope,
2007 June 24.93 UT, with Lanning 386 in quiescence.  The scale and orientation are
as indicated.  A coordinate solution for this image based on numerous UCAC-2 stars 
\citep{zacharias04}
puts Lanning 386 at $\alpha = 21^{\rm h} 08^{\rm m} 33^{\rm s}.97, +39^{\circ} 05' 35''.3$,
referred to the ICRS (equinox 2000).}
%are labeled as in Table \ref{cmptable}. North is up and east to the left in the figure.}
%A POSS2 blue image covering 15'x15' field-of-view with Lanning 386 marked. The comparison stars
%are labeled as in Table \ref{cmptable}. North is up and east to the left in the figure.}
\label{finder_wide}
\end{figure}

%
%\begin{figure}
%\plotone{f4.eps}
%\caption{Power spectrum for $3.0 < f < 20.0$ cycles/day from the CLEANest algorithm \citep{foster1995}.
%The arrows mark the orbital period at 6.095 cycles/day and harmonic frequencies
%at 12.19 and 3.084 cycles/day.}
%\label{power}
%\end{figure}
%
%\begin{figure}
%\plotone{f10.eps}
%\caption{Average of four spectra taken near phase 0.75 showing inverse P-Cygni absorption for
%all the \ion{He}{1} lines and for Balmer lines starting with H$\delta$. The feature at 517.5~nm
%identified as \ion{Mg}{1} is present in the spectrum at this phase.}
%\label{spec2plot}
%\end{figure}
%

\begin{thebibliography}{}


\bibitem[Baraffe \& Kolb(2000)]{baraffe} Baraffe, I., \& Kolb, U.\ 2000, \mnras, 318, 354

\bibitem[Bessell(1990)]{bessell} Bessell, M. S. 1990, \pasp, 102, 1181

\bibitem[Beuermann(2006)]{beuermann06} Beuermann, K.\ 2006, \aap, 460, 78

\bibitem[Bruch \& Engel(1994)]{bruch94} Bruch, A., \& Engel, 
A.\ 1994, \aaps, 104, 79 

%\bibitem[Casares et al.(1996)]{casares96} Casares, J., Martinez-Pais, I.G., Marsh, T. R., Charles, P.A., Lazaro, C. 1996, MNRAS, 278, 219

\bibitem[Downes  et al.(2001)]{downes2001} Downes, G. et al. 2001, PASP, 113, 764

\bibitem[Eracleous et al.(2002)]{eracleous2002} Eracleous, M. et al. 2002, PASP, 114, 207E

% \bibitem[Foster (1995)]{foster1995} Foster, G. 1995, AJ, 109, 1889

% \bibitem[Garnavich \& Szkody(1992)]{garnavich1992}Garnavich, P.M. and Szkody, P. 1992, AAVSOJ, 21, 81

\bibitem[Groot et al.(2000)]{groot2000} Groot, P. J., Rutten, R. G. M., and van Paradijs, J. 2000, New Astro. Rev., 44, 137

\bibitem[Hellier et al.(2000)]{hellier00} Hellier, C., Kemp, J., 
Naylor, T., Bateson, F.~M., Jones, A., Overbeek, D., Stubbings, R., \& 
Mukai, K.\ 2000, \mnras, 313, 703    % EX Hya outburst

\bibitem[Hellier et al.(1989)]{hellier89} Hellier, C., Mason, 
.~O., Smale, A.~P., Corbet, R.~H.~D., O'Donoghue, D., Barrett, P.~E., \& 
Warner, B.\ 1989, \mnras, 238, 1107  % EX Hya outburst spectra

\bibitem[Hellier et al.(1997)]{hellier97} Hellier, C., Mukai, K., 
\& Beardmore, A.~P.\ 1997, \mnras, 292, 397 

\bibitem[Hessman et al.(1984)]{hessman84} Hessman, F.~V., 
Robinson, E.~L., Nather, R.~E., \& Zhang, E.-H.\ 1984, \apj, 286, 747 

%\bibitem[Hoard et al.(2003)]{hoard03} Hoard, D.~W., Szkody, P., 
%Froning, C.~S., Long, K.~S., \& Knigge, C.\ 2003, \aj, 126, 2473 

%\bibitem[Honeycutt, Livio, \& Robertson(1993)]{honeycutt1993}Honeycutt, R. K., Livio, M., \& Robertson, J. W. 1993, PASP, 105 922

\bibitem[Honeycutt \& Kafka(2004)]{honeycutt04} Honeycutt, R.~K., 
\& Kafka, S.\ 2004, \aj, 128, 1279  % many VY Scl light curves.

\bibitem[Horne et al.(1982)]{Horne1982}Horne, K., Lanning, H.H., and Gomer, R.H. 1982, Ap. J. 252, 681

\bibitem[Landolt(1992)]{landolt92} Landolt, A. U. 1992, AJ, 104, 340

\bibitem[Lanning(1973)]{lanning1973} Lanning, H.H. 1973, PASP, 85 70

\bibitem[Lanning \& Meakes(1998)]{lanning1998} Lanning, H.H. and Meakes, M. 1998, PASP, 110, 586

\bibitem[Lanning \& Meakes(2000)]{lanning2000} Lanning, H.H. and Meakes, M. 2000, PASP, 112, 251

%\bibitem[Leach et al.(1999)]{leach1999}Leach, R., Hessman, F. V., King, A.R., Stehle, R., \& Mattei, J. 1999, MNRAS, 305, 225

\bibitem[Liu \& Hu(2000)]{liu00} Liu, W., \& Hu, J.~Y.\ 2000, \apjs, 128, 387 

%\bibitem[Martinez-Pais et al.(1999)]{martinez99} Martinez-Pais, I.G., Rodriguea-Gil, P., Casares, J. 1999, MNRAS, 305, 661

%\bibitem[Norsworthy(1991)]{norsworthy1991} Norsworthy, J.E. 1991, AJ, 102, 272

%\bibitem[Patterson et al.(2002)]{patterson02} Patterson, J., Fenton, W. H., Thorstensen, J.R., et al. 2002, PASP, 114, 1364

\bibitem[Patterson(1994)]{patterson94} Patterson, J.\ 1994, \pasp, 
106, 209 

%\bibitem[Patterson et al.(2002)]{patterson02} Patterson, J., et 
%al.\ 2002, \pasp, 114, 1364   % SWs may be magnetic

\bibitem[Rodr{\'{\i}}guez-Gil et al.(2004)]{rodriguez04} 
Rodr{\'{\i}}guez-Gil, P., G{\"a}nsicke, B.~T., Barwig, H., Hagen, H.-J., \& 
Engels, D.\ 2004, \aap, 424, 647 

%\bibitem[Rodriguez-Gil et al.(2007)]{rodriguez2007}Rodriguez-Gil, P., et al. 2007, MNRAS, 377, 1747 

\bibitem[Schlegel et al.(1998)]{schlegel} Schlegel, D.~J.,  
Finkbeiner, D.~P., \& Davis, M.\ 1998, \apj, 500, 525

\bibitem[Schmidtobreick(2003)]{schmidtobreick2003} Schmidtobreick, L., Tappert, C., Bianchini, A., and Mennickent, R. E. 2003, A\&A, 410, 943

\bibitem[Schneider \& Young(1980)]{schyo} Schneider, D.~P.~\&
Young, P.\ 1980, \apj, 238, 946

\bibitem[Schneider et al.(1981)]{schneider81} Schneider, D.~P., 
Young, P., \& Shectman, S.~A.\ 1981, \apj, 245, 644 

\bibitem[Shafter(1983)]{shaf} Shafter, A.~W.\ 1983, \apj, 267, 222

\bibitem[Shafter et al.(1983)]{shafter1983} Shafter A.W., Lanning, H.H., Ulrich R.K 1983, PASP, 95, 206

%\bibitem[Shafter et al.(1985)]{shafter85} Shafter, A.~W., Szkody, 
%P., Liebert, J., Penning, W.~R., Bond, H.~E., \& Grauer, A.~D.\ 1985, \apj, 
%290, 707 

\bibitem[Sheets et al.(2007)]{sheets07} Sheets, H.~A., 
Thorstensen, J.~R., Peters, C.~J., Kapusta, A.~B., \& Taylor, C.~J.\ 2007, \pasp, 119, 494

%\bibitem[{\v S}imon(2000)]{simon00} {\v S}imon, V.\ 2000, \aap, 
%360, 627  % DO Dra outburst stats

\bibitem[Stetson(1987)]{stetsondao} Stetson, P.~B.\ 1987, \pasp,
99, 191

\bibitem[Thorstensen et al.(1991)]{thorstensen1991} Thorstensen, J.R., Ringwald, F. A., 
Wade, R. A., Schmidt, G.D., and Norsworthy, J.E. 1991, AJ, 102, 272

\bibitem[Thorstensen et al.(1998)]{thorstensen98} Thorstensen, J.~R., 
Taylor, C.~J., \& Kemp, J.\ 1998, \pasp, 110, 1405 

\bibitem[Warner(1995)]{warner} Warner, B.\ 1995, Cambridge
Astrophysics Series, Cambridge, New York: Cambridge University Press,
|c1995

\bibitem[Zacharias et al.(2004)]{zacharias04} Zacharias, N., Urban, 
S.~E., Zacharias, M.~I., Wycoff, G.~L., Hall, D.~M., Monet, D.~G., \& 

\end{thebibliography}
\end{document}